\documentclass[10pt]{article}
\usepackage{epsfig}
\usepackage{alltt}
\usepackage{multirow}
\usepackage{xcolor}

\newcommand{\lcp}{\mbox{\it lcp}}
\newcommand{\lcpp}{\mbox{\it lcp3}}

\begin{document}

\title{Merging Sorted Lists of Similar Strings}
\author{
        Gene Myers \\
        Okinawa Institute of Science and Technology \\
        Okinawa 904-0495, JAPAN \\
        {\it and} \\
        MPI for Molecular Cell Biology and Genetics \\
        01307 Dresden, GERMANY \\
        $\mathtt{email: gene.myers@gmail.com}$
}
\date{}
\maketitle
\vspace{-.25in}
\begin{abstract}
Merging $T$ sorted, non-redundant lists containing $M$ elements into a single sorted, non-redundant result of size $N \ge M/T$ is a classic problem
typically solved practically in $O(M \log T)$ time with a priority-queue data structure the most basic of which is the
simple {\it heap}.
We revisit this problem in the situation where the list elements are {\it strings} and
the lists contain many {\it identical or nearly identical elements}.
By keeping simple auxiliary information with each heap node, we devise
an $O(M \log T+S)$ worst-case method that performs no more character comparisons than the sum of the lengths of
all the strings $S$, and another $O(M \log (T/ \bar e)+S)$ method that becomes progressively more efficient as a function of the fraction of equal elements $\bar e = M/N$
between input lists, reaching linear time when the lists are all identical.
The methods perform favorably in practice versus an alternate formulation based on a trie.
\end{abstract}

\section{Introduction \& Summary}

Producing a sorted list, possibly with duplicate elements removed, from a collection of $T$
sorted input lists is a classic problem \cite{Knuth}.  Moreover, with today's massive data sets,
where an in-memory sort would require an excessively large memory, this problem gains in importance as a
component of an external, disk-based sort.
Our motivating example is modern DNA sequencing projects that involve anywhere from
100 billion to 5 trillion DNA bases of data in the form of sequencing reads that are conceptually
strings over the 4-letter alphabet A, C, G, T \cite{VGP}.  In particular, the problem of producing a sorted table
of all the $k$-mers (substrings of length exactly $k$) and their counts has been the focus of much
study and is used in many analysis methods for these data sets \cite{KMC3,PBJelly,FastK}.

Priority queue implementations such as a heap, take $O(\log T)$ to extract the next minimum and insert
its replacement, giving an $O(M \log T)$ merge time where $M$ is the sum of the lengths of the input lists \cite{Corman}.
However when the domain of the merge is strings, as opposed to say integers, then one must consider
the time taken for each of the $O(\log T)$ string comparisons, which is not $O(1)$ but
conceptually the average length of the longest common prefix ($\lcp$) between all the compared strings.
For example, this is $O(\log_\Sigma M)$ in the ``{\it Uniform Scenario}" where the characters of the strings are chosen with equal
probability over an alphabet of size $\Sigma$.
But it can be much worse, for example, when merging lists of say $21$-mers each obtained from a portion
of a 40X coverage DNA sequencing data set, where many strings are identical.

In the worst case, one can only assert that the time to merge the list of strings is $O(S \log T)$ where $S$ is the total number of
characters in the input lists, e.g. $Mk$ for lists of $k$-mers.
Assuming the Uniform Scenario, one can more accurately characterize the efficiency as
$O(M \log T \log M)$ {\it expected} time.
In this paper we present a method that
is guaranteed to take $O(M \log T + S)$ time by modifying the heap data structure so that
the {\it amortized} time spent on comparing the characters of any string while it is in the heap is never
more than its length.  Moreover, in the Uniform Scenario, the efficiency is $O(M (\log T + \log M))$
expected time.
We call such a modified heap a {\it string heap}.

It is further true in the case of DNA sequencing data sets, that often the number of elements $N$ in the merged list is much smaller
than $M$ when duplicate elements are removed.  Specifically, $N$ can be as small as $M/T$ assuming the input
lists themselves do not contain equal elements.
With another modification to a heap, not specific to strings {\it per se}, we will achieve here an algorithm
that takes $O(M \log (T/\bar e))$ time where $\bar e = M/N$ is the average number of distinct input lists a given element
is in.  So when all the input elements are unique the time is as usual $O(M\log T)$ but as $\bar e$ increase less
time is taken, reaching $O(M)$ when all the input lists are identical, that is, $\bar e = T$.
We call such a modified heap a {\it collision heap}.  We show it can easily be combined with a string heap to
give an $O(M \log (T/\bar e) + S)$ algorithm for string merging.

While the focus of this paper is on modifying a heap to support string elements, an orthogonal
approach to realizing a priority queue (PQ) of strings appeared in a comprehensive paper by Thorup (\cite{Thorup})
that is primarily focused on integer PQs, but which in Section 6 uses a trie {\cite{Fredkin} to merge strings of, potentially large, integers
in $O(M \log \log T+S)$ time.  In bioinformatics, strings are generally over alphabets of small size $\Sigma$, e.g. 4 for DNA, so taking Thorup's
algorithm, but replacing the general integer priority queue with  van Emde Boas small integer PQs {\cite{Boas} over domain $\Sigma$,
one obtains an $O(M \log \log \Sigma + S)$ time algorithm.
The implementation of either of these methods encounters rather larger overheads compared to simply realizing the basic approach of Thorup's algorithm with a compact trie with $\Sigma$-element arrays for the out-edges.  Moreover, because adding to the trie then becomes linear,
the complexity of this simplified approach is $O(N \Sigma + S)$.
Given limited values of $\Sigma$, e.g. say up to 20 for protein sequences, the trie approach is very
competitive, especially for the cases where $N$ is significantly smaller than $M$.

We implemented programs to merge files of sorted strings using a regular heap, a string heap, a collision heap, a combination of the
string and collision heap, and a simple compact trie and performed timing experiments on both simulated and real DNA sequencing
$k$-mer data sets to determine their performance relative to each other.  The codes are available at {\tt github.com/thegenemyers/HEAPS}.  Amongst the heap-based algorithms, the string heap proves
superior as the average $\lcp$ between consecutive output strings increases, and the collision heap proves superior as the collision ratio $\bar e$ increases.
Also, the combination heap tracked the behavior of whichever of the string or collision heap proves superior, but at an
overhead of roughly 5\%.  Against the trie approach, the string heap is faster in the uniform scenario until $T$ becomes quite large,
e.g. 256 in our experiments.  In scenarios where $N \ll M$ due to a {\it uniform} collision rate the trie proved fastest save for small values of $T$.
For real data sets where the collision rate is highly variable, the collision and combination heaps actually gave the best times.
In short, the new heap methods are of both theoretical and practical interest.

\section{Preliminaries: Definitions and a Short Recap of Heaps}

Consider $T$ sorted lists of strings $S_t = s_1^t, s_2^t, \ldots s_{N_t}^t$ of lengths $N_t$.
We assume that the elements are distinct, i.e. $s_j^t < s_{j+1}^t$, and let $s_j^t = a_0^{j^t} a_1^{j^t} \ldots a_{n_j^t-1}^{j^t}$.
Note carefully, that the first character of a string is at index 0.
The problem is to produce a single sorted list $R = r_1, r_2, \ldots r_N$ of length $N$ with any duplicates {\it between}
the lists removed.
That is, while each input list has unique strings, the same string, can occur in up to $T$ different lists.
Let $e_i \in [1,T]$ be the number of different queues the string $r_i$ occurs in.
Letting $M = \sum_{t=1}^T N_t$ be the sum of the lengths of the input lists,
note that $N$ is in the range $[M/T , M]$.

A $T$ element heap is a complete binary tree of $T$ nodes containing or referring to domain values to be prioritized.
A heap further has the {\it heap property} when for every node, the domain values of its children are greater than its 
domain value.
A heap can be very simply implemented as an array $H$[$1..T$] where $H[i]$ is the datum for node $i$, its
left child is $2i$, and its right child is $2i+1$ (if they exist, i.e. are $\le T$).

In the case of merging $T$ input lists, we will let each heap node contain the index $t \in [1,T]$ of an input list and another
array, $V[0..T]$ will contain the current value for that list in $V[t]$ (the role of $V[0]$ is discussed in the next paragraph).
If all the nodes greater than $i$ have the heap property, then recall that the following simple routine {\it Heapify(i,x,t)} will add
the value $x$ from list $t$ to the heap guaranteeing that $H$ has the heap property for all nodes greater than $i\!-\!1$.
The routine takes time proportional to at most the height of $i$ in the heap which is $O(\log T)$ for all $i$.  Pseudo-code
for $Heapify$ is shown at left in Figure 1.

Let $S[t]$ denote the $t^{th}$ sorted input list and assume it operates as a one-sided queue where one can
{\it Pop} the next element from the list and ask if the queue is {\it Empty}.   We will also assume that {\it INFINITY}
is an infinitely large string value greater than all those encountered as input and when a list is exhausted place
this value at $V[0]$ so that $H[1]$ becomes 0 when {\it all} the lists are exhausted.  Then a complete pseudo-code for the
basic priority queue approach to merging $T$ sorted lists {\it while removing duplicate values} is shown at right in Figure 1.

\begin{figure}[h]
\scriptsize
\begin{verbatim}
        int    T                                                          domain_list S[1..T]    
        int    H[1..T]
        domain V[0..T]                                                 1. for t = T downto 1 do
                                                                       2.   Heapify(t,Pop(S[t]),t)
        Heapify(i,x,t)
        { c = i                                                        3. last = INFINITY
          V[t] = x                                                     4. while (t = H[1]) > 0 do
          while (u=2c) <= T do                                         5.   { x = V[t]
            { if u < T and V[H[u+1]] < V[H[u]] then                    6.     if x > last then
                u = u+1                                                7.       output (last=x)
              if x <= V[H[u]] then                                     8.     if Empty(S[t]) then
                break                                                  9.       Heapify(1,INFINITY,0)
              H[c] = H[u]                                             10.     else
              c = u                                                   11.       Heapify(1,Pop(S[t]),t)
            }                                                               }
          H[c] = t                     
        }
\end{verbatim}
\normalsize
\caption{The {\it Heapify} routine (left) and the overall merge algorithm (right).}
\end{figure}

In lines 1 and 2, the first element of each list is {\it Pop}'d and placed in the heap in reverse order of the nodes so that the
entire heap has the heap property upon completion.
The total time taken for this setup is $O(T)$ as the sum of the heights of the nodes in a complete
binary tree is of this order.
Then in the while-loop of line 4, the list $t$ with the next smallest element is $H[1]$ and if this value is not zero,
indicating the exhaustion of all the queues, then the element $x = V[t]$ is processed in the loop body.
If the value $x$ is not a duplicate of the last element output then it is output (lines 5 - 7).
If list $t$ is not empty then its next element replaces the element just output and the heap property is restored at node 1 (lines 8 \& 11).  Otherwise the element is replaced with the largest possible value {\it INFINITY} (line 9) in "queue" $0$ so that when all $T$ lists are exhausted
the extraction of 0 as the queue index marks the end of the merge.

Given that {\it Heapify} takes $O(\log T)$ time and an input element is processed with each iteration of the loop, the
algorithm clearly operates in $O(M log T)$ time assuming domain comparisons are $O(1)$.  As discussed in the introduction
this assumption is not necessarily true when the values are strings and we address this in the next section.

\section{The String Heap}

The idea for a string heap is very simple, namely, for each node also record and keep current the length
of the longest common prefix between the string at a node and the string at its parent (except for the root).
Let $\lcp(u,v)$ be the longest common prefix between strings $u$ and $v$.  Then more formally, a string heap
also maintains a third array $P[1..T]$ such that $P[i] = \lcp(V[H[i]],V[H[\lfloor i/2\rfloor]])$ for $i > 1$.
The interesting and complex part of this extension is maintaining this property during the induction of {\it Heapify}
and using it to accelerate the comparison of string values by limiting the number of character comparisons involved.

Intuitively, {\it Heapify(i,x,t)} traverses the maximal left most path starting at $i$, all of whose elements are less than $x$
and not more than their siblings until a node $c$ is reached that is either a leaf or for which all its children are not less than $x$.
The values along this path are shifted up to the node above during each iteration until $x$ is placed at node $c$ at the last.
To help argue the induction to follow, it conceptually simplifies matters to think of $x$ as being explicitly placed at node $c$
(i.e. $H[c] = t$) as the algorithm descends from node $i$ to the final placement of $x$.  From this viewpoint, at the start of each iteration
of the loop of {\it Heapify}, the heap satisfies the heap property at every node in the subtree rooted at $i$ {\it except} $c$
where $x$ conceptually currently resides.  For the array $P$ realizing a string heap, the loop invariant is that $P$ is correct
except possibly at nodes $2c$ and $2c+1$ as $x$ has just been placed at their parent node $c$.  Our goal is to maintain
this invariant through the next iteration of the loop where either $x$ is found to be not greater than the children of $c$
and the loop exits, or the algorithm descends to one of the children of $c$ swapping $x$ with the child's value.

If $c > i$ then in the previous iteration the value that was at $c$ is now at $\lfloor c/2 \rfloor$ having been exchanged
with $x$ as it has a smaller value.
Let $o = V[H[\lfloor c/2 \rfloor]] < x$ be this value and also let $v_l = V[H[2c]]$ and $v_r = V[H[2c+1]]$
be the strings currently at the left and right children of $c$.
Then it must be that $P[2c] = \lcp(o,v_l)$ and
$P[2c+1] = \lcp(o,v_r)$ as these values are unchanged since the previous iteration when $o$ was at node $c$.
Let $p_l$ and $p_r$ denote these
values, respectively, and further let $p = P[c] = \lcp(o,x)$.
A final subtlety is that when $x$ is conceptually placed at $c$, $P[c]$ is only
conceptually set to this value which is actually kept in an explicit program variable $p$.

To start the induction when $c = i = 1$, observe that it will be the case that $P[2]$ and $P[3]$ will have the
value $\lcp(o,H[V[2]])$ and $\lcp(o,H[V[3]])$ where $o$ is the value that was just extracted from the root of the
heap and which $x$ is now about to replace.  So in order to get started we need to set $p$ to
$\lcp(V[H[1]],x)$.  In the context of list merging, $o$ was the previous value on the queue $H[1]$ that
$x$ is now replacing, and so in this context we know that $o \le x$, and so in fact with this choice of
$p$ we have exactly the inductive invariant described in the previous paragraph.  Indeed one could
imagine the $o$ is at the virtual father of the root $1$.  The reader should observe that if $o > x$ then
getting the induction started also requires readjusting $P[c]$ and $P[2c+1]$ downward to $p$
if they happen to be larger than $p$.   In this case, we can no longer place a bound on the total number
of character comparison made during the operation of the heap, but it will still operate correctly.  Formally, our solution
is said to be restricted to {\it monotone} problems, that is, the value replacing a just extracted value must be monotone increasing.
Finally, when $i > 0$, which only occurs when one first constructs the heap, one simply assumes that $o$ is the empty string,
$\epsilon$, so that $P[i] = P[2i] = P[2i+1] = 0$.

To facilitate a simpler logic around the comparison of strings, we will assume that every string
ends with a special terminating character $\$$ that is less than any ordinary character (e.g. 0 for C-strings).
With this convention, finding the $\lcp$ of two strings $x$ and $y$ is simply a matter of finding the first index $\rho$ for
which the strings have unequal characters or both are $\$$.  Moreover note that $x < y$ iff $x[\rho] < y[\rho]$ (recalling
that string indexing begins at 0).

We now proceed to analyze the numerous cases that arise in terms of the relationships between the
quantities $p$, $p_l$, and $p_r$.  To further simplify matters observe that the treatment of the left and
right children of $c$ is symmetric, so we only consider the left case, $p_l \le p_r$, in the enumeration below knowing
that the right case, $p_r < p_l$, is handled simply by exchanging the roles of left and right. 
Furthermore, we repeatedly use the logic that if $\lcp(x,s) < \lcp(s,y)$ then $\lcp(x,y) = \lcp(x,s)$ and $x < s$ iff $x < y$.

\vspace{8pt}
{\bf Case 1:} $p_r < p_l$ and $p < p_l$:

\vspace{3pt}
\hspace*{3pt}
\begin{minipage}{.9\textwidth}
By the case condition $\lcp(x,o) = p < p_l = \lcp(o,v_l)$ and since we know $x > o$ we can conclude
that $x > v_l$ and $\lcp(x,v_l) = \lcp(x,o) = p$.
Similarly $\lcp(v_r,o) = p_r < p_l = < \lcp(o,v_l)$ and we know $v_r \ge o$ allowing us to conclude
that $v_r > v_l$ and $\lcp(v_r,v_l) = \lcp(v_r,o) = p_r$.
So $v_l$ is the smallest of $x$, $v_l$, and $v_r$ implying that the loop should descend to $2c$ with
$v_l$ being placed at $c$.
Moreover, $P[c]$ should be set to $p_l$, while $p$ and $P[2c+1]$ can remain unchanged having already the
correct values for the next iteration.
\end{minipage}

\vspace{8pt}
{\bf Case 2:} $p_r \le p_l$ and $p > p_l$:

\vspace{3pt}
\hspace*{3pt}
\begin{minipage}{.9\textwidth}
By the case condition $\lcp(v_l,o) = p_l < p = \lcp(o,x)$ and since we know $v_l \ge o$ we can conclude
that $x < v_l$ and $\lcp(v_l,x) = \lcp(v_l,o)$.
Similarly $\lcp(v_r,o) = p_r \le p_l < p = \lcp(o,x)$ and we know $v_r \ge o$ allowing us to conclude
that $x < v_r$ and $\lcp(v_r,x) = \lcp(v_r,o)$.
So the loop can terminate with $x$ being placed at node $c$.
Moreover, $P[2c]$ and $P[2c+1]$ remain unchanged having yet the correct values.
\end{minipage}

\vspace{8pt}
{\bf Case 3:} $p_r < p_l$ and $p = p_l$:

\vspace{4pt}
\hspace*{3pt}
\begin{minipage}{.9\textwidth}
First compute $p_x = p + \lcp(v_l+p,x+p)$ where $s+j$ is the suffix of string $s$ beginning at
position $j$.  Clearly $p_x = \lcp(v_l,x)$ and if ${v_l}[p_x] < x[p_x]$ then $v_l < x$, otherwise $v_l \ge x$.
We have two subcases:
\end{minipage}

\vspace{8pt}
\hspace*{3pt}
{\bf Subcase 3a:} $v_l[p_x] < x[p_x]$:

\vspace{4pt}
\hspace*{9pt}
\begin{minipage}{.9\textwidth}
The condition $p_r < p_l$ implies $\lcp(v_r,o) <  \lcp(o,v_l)$ and we know $v_r \ge o$
allowing us to conclude that $v_r > v_l$ and $\lcp(v_r,v_l) = \lcp(v_r,o)$.
Thus $v_l$ is smaller than both $x$ and $v_r$.
So the loop should descend to $2c$ with $v_l$ being placed at $c$.
Therefore, $P[c]$ should be set to $p_l$ and $p$ to $p_x$, while $P[2c+1]$ has the correct value.
\end{minipage}

\vspace{8pt}
\hspace*{3pt}
{\bf Subcase 3b:} $v_l[p_x] \ge x[p_x]$:

\vspace{4pt}
\hspace*{9pt}
\begin{minipage}{.9\textwidth}
By the case conditions we know $p_r < p$ implying $\lcp(v_r,o) <  \lcp(o,x)$ and we know $v_r \ge o$
allowing us to conclude that $v_r > x$ and $\lcp(v_r,x) = \lcp(v_r,o)$.
Thus $x$ is not less than both $v_l$ and $v_r$.
So the loop can terminate with $x$ being placed at node $c$.
While $P[2c+1]$ remains correct, $P[2c]$ needs to be updated to $p_x$.
\end{minipage}

\newpage
\vspace{8pt}
{\bf Case 4:} $p_r = p_l$ and $p < p_l$:

\vspace{3pt}
\hspace*{3pt}
\begin{minipage}{.9\textwidth}
Compute $p_x = p_l + \lcp(v_l+p_l,v_r+p_l)$ which is clearly $\lcp(v_l,v_r)$.
If $v_r[p_x] < v_l[p_x]$ then $v_r < v_l$, otherwise $v_r \ge v_l$.
WLOG let's assume $v_l \le v_r$, as the case $v_r < v_l$ is symmetric.
As in cases before $p < p_l$ and $x > o$ allow us to surmise that
$x > v_l$ and $\lcp(x,v_l) = \lcp(x,o)$.
Therefore $v_l$ is smaller than $x$ and not larger than $v_r$, implying that
the loop should descend to $2c$ with $v_l$ being placed at $c$.
Therefore, $P[c]$ should be set to $p_l$, $P[2c+1]$ to $p_x$, while
$p$ continues to have the correct value.
\end{minipage}

\begin{figure}
\scriptsize
\begin{verbatim}
Heapify(int i, string x, int t)                                           int LCP2(string x, string y, int n)
{ c = i                                                                   { while true do
  p = LCP2(H[V[i]],x,0)                                                          { (a,b) = (x[n],y[n])
  while (l = 2c) <= T do                                                         if a != b or a == $ then
    { (hl,pl) = (H[l],P[l])                                                        return n
      if l < T then                                                              n += 1        
        (hr,pr) = (H[l+1],P[l+1])                                              }
      else                                                                }
        pr = -1
      if pr < pl then
        { if p < pl then                             # Case 1L            int LCP3(string x, string y, string z, int n)
            (H[c],P[c],c) = (hl,pl,l)                                     { while true do
          else if p > pl then                        # Case 2L                { (a,b,c) = (x[n],y[n],z[n])
            break                                                               if a != b or a != c or a == $ then
          else                                                                    return n
            { vl = V[hl]                                                        n += 1
              px = LCP2(vl,x,pl)                                              }
              if vP[px] < x[px] then                 # Case 3La           }
                (H[c],P[c],p,c) = (hl,pl,px,l)
              else                                   # Case 3Lb
                { P[l] = px
                  break
                }
            }
        }
      else if pr > pl then
        { # Case 1R, 2R, 3Ra, 3Rb
          ...
        }
      else if p > pl then                           # Case 2
        break
      else
        { (vl,vr) = (V[hl],V[hr])      
          if p < pl then                            # Case 4
            { px = LCP2(vr,vl,pl)
              if (vl[px] <= vr[px])
                (H[c],P[c],P[l+1],c) = (hl,pl,px,l)
              else
                (H[c],P[c],P[l],c) = (hr,pr,px,l+1)
            }
          else                                      # Case 5
            { px = LCP3(vl,vr,x,p)
              if vr[px] > vl[px] then
                { if x[px] > vl[px] then            # Case 5.1L
                    (H[c],P[c],P[l+1],p,c) = (hl,pl,px,px,l)
                  else if x[px] < vl[px] then       # Case 5.2L
                    { P[l] = P[l+1] = px
                      break
                    }
                  else
                    { py = LCP2(vl,x,px)
                      if vl[py] < x[py] then        # Case 5.3La
                        (H[c],P[c],P[l+1],p,c) = (hl,pl,px,py,l)
                      else                          # Case 5.3Lb
                        { (P[l],P[l+1]) = (py,px)
                          break
                        }
                    }
                }
              else if vr[px] < vl[px] then
                { # Case 5.1R, 5.2R, 5.3Ra, 5.3Rb
                  ...
                }
              else if x[px] <= vl[py] then          # Case 5.2
                { P[l] = P[l+1] = px
                  break
                }
              else                                  # Case 5.4
                { py = LCP2(vl,vr,px)
                  if vl[py] < vr[py] then
                    (H[c],P[c],P[l+1],p,c) = (hl,pl,py,px,l)
                  else
                    (H[c],P[c],P[l],p,c) = (hr,pr,py,px,l+1)
                }
            }
        }
    }
  (H[c],P[c],V[t]) = (t,p,x)
}
\end{verbatim}
\normalsize
\caption{The {\it Heapify} algorithm for a string heap.}
\end{figure}

\vspace{6pt}
{\bf Case 5:} $p_r = p_l = p$:

\vspace{3pt}
\hspace*{3pt}
\begin{minipage}{.9\textwidth}
First compute $p_x = p + \lcpp(v_l+p,v_r+p,x+p)$ where $\lcpp$ is the 3-way common prefix, and
this by the case conditions is clearly equal to $\lcpp(v_l,v_r,x)$.
There now arise numerous subcases based on the relationships between
$x[p_x]$, $v_l[p_x]$, and $v_r[p_x]$ in direct analogy to the subcases based on
the relationships between $p$, $p_l$, and $p_r$, so we will number these 5.1, 5.2, and so on:
\end{minipage}

\vspace{6pt}
\hspace*{3pt}
{\bf Subcase 5.1:} $v_r[p_x] > v_l[p_x]$ and $x[p_x] > v_l[p_x]$:

\vspace{3pt}
\hspace*{9pt}
\begin{minipage}{.9\textwidth}
The case conditions imply $x > v_l$ and $v_r > v_l$ and $\lcp(v_l,x) = \lcp(v_l,v_r) = p_x$.
So $v_l$ is the smallest of $x$, $v_l$, and $v_r$ implying that the loop should descend to $2c$ with
$v_l$ being placed at $c$.
So $P[c]$ should be set to $p_l$, while $p$ and $P[2c+1]$ are now clearly $p_x$.
\end{minipage}

\vspace{6pt}
\hspace*{3pt}
{\bf Subcase 5.2:} $v_r[p_x] \ge v_l[p_x]$ and $x[p_x] \le v_l[p_x]$:

\vspace{3pt}
\hspace*{9pt}
\begin{minipage}{.9\textwidth}
In this subcase, clearly $x$ is not more than both $v_l$ and $v_r$ and $\lcp(v_l,x) = \lcp(v_r,x) = p_x$.
So the loop should terminate and both $P[2c]$ and $P[2c+1]$ should be updated to $p_x$.
\end{minipage}

\vspace{6pt}
\hspace*{3pt}
{\bf Subcase 5.3:} $v_r[p_x] > v_l[p_x]$ and $x[p_x] = v_l[p_x]$:

\vspace{3pt}
\hspace*{9pt}
\begin{minipage}{.9\textwidth}
First compute $p_y = p_x + \lcp(v_l+p_x,x+p_x)$ which is clearly $\lcp(v_l,x)$
note that the conditions to this point imply $\lcp(v_l,v_r)$ = $\lcp(x,v_r) = p_x$.
\end{minipage}

\vspace{6pt}
\hspace*{9pt}
{\bf Subcase 5.3a:} $v_l[p_y] < x[p_y]$:

\vspace{3pt}
\hspace*{15pt}
\begin{minipage}{.9\textwidth}
So $v_l$ is the smaller than $x$ and $v_r$ implying the loop should descend to $2c$ with
$v_l$ being placed at $c$.  So $P[c]$ should be set to $p_l$ and the correct new values
for $p$ and $P[2c+1]$ are $p_y$ and $p_x$, respectively.
\end{minipage}

\vspace{6pt}
\hspace*{9pt}
{\bf Subcase 5.3b:} $v_l[p_y] \ge x[p_y]$:

\vspace{3pt}
\hspace*{15pt}
\begin{minipage}{.9\textwidth}
So $x$ is not smaller than $v_l$ and $v_r$ implying the loop can terminate at $c$,
where $P[2c]$ and $P[2c+1]$ are updates to $p_y$ and $p_x$, respectively.
\end{minipage}

\vspace{6pt}
\hspace*{3pt}
{\bf Subcase 5.4:} $v_r[p_x] = v_l[p_x]$ and $x[p_x] > v_l[p_x]$:

\vspace{3pt}
\hspace*{9pt}
\begin{minipage}{.9\textwidth}
Compute $p_y = p_x + \lcp(v_l+p_x,v_r+p_x)$ which is clearly $\lcp(v_l,v_r)$.
If $v_r[p_y] < v_l[p_y]$ then $v_r < v_l$, otherwise $v_r \ge v_l$.
WLOG let's assume $v_l \le v_r$, as the case $v_r < v_l$ is symmetric.
By the case conditions $x > v_l$ and $\lcp(x,v_l) = p_x$.
Therefore $v_l$ is smaller than $x$ and not larger than $v_r$, implying that
the loop should descend to $2c$ with $v_l$ being placed at $c$.
Therefore, $P[c]$ should be set to $p_l$, $P[2c+1]$ to $p_y$, and $p$ to $p_x$.
\end{minipage}

\newcommand{\LCP}{\mbox{\tt LCP2}}
\newcommand{\LCPP}{\mbox{\tt LCP3}}

\vspace{6pt}
Figure 2 presents the complete algorithm for the string version of {\it Heapify} embodying
the case analysis above so that the $P$-array values are correctly maintained.
Note carefully, that the $\lcp$ information in the $P$-array is used both to determine
the relative values of the heap elements and hence direct the path that ${\it Heapify}$
takes to insert a new element $x$, but further also saves time on the number of character
comparisons performed by only computing new $\lcp$'s in terms of an initial $\lcp$-offset that is
common to {\it all} of the arguments to $\LCP$ or $\LCPP$.

As regards complexity, the algorithm for {\it Heapify} takes $O(\log T)$ time {\bf plus} the time
spent in $\LCP$ or  $\LCPP$ for character comparisons.  Note carefully the code assumes
that we are merging {\it sorted} string lists, so the value of $x$ is not less than the value of the
previous element $o = V[H[1]]$ on the same queue $H[1]$.  We make an amortization argument
to bound the total number of character comparisons as follows:

First, observe that every string value has an $\lcp$-value associated with it, namely,
for the string $V[H[i]]$ it is $P[i]$ and it represents the number of character comparisons "charged" to
its string.
Examination of the case conditions reveals that when $\LCP$ or $\LCPP$ is called, all the string
arguments have the same {\it lcp-value} at the time of the call.
Afterwords, all but one of the arguments will have its $\lcp$-value increased to the returned value, 
effectively charging the comparisons of the $\lcp$ call to those arguments (NB: for $\LCPP$ {\bf two} comparisons
per $\lcp$ increment are made).  The total time taken then over the course of the merge is the sum of the
maximum $\lcp$-value of every string that passes through the heap.  Since the $\lcp$-value of each string is never
more than the length of the string, we have our $O(S)$ bound on the total number of character comparisons.

We can more accurately characterize the number of comparisons with the observation that the maximum
$\lcp$-value that each string reaches when it is extracted from the root of the heap is its $\lcp$ with the
string value extracted just before it.  To see this simply review WLOG the logic involved in a value moving
from node $2$ to node $1$ where, in all relevant cases, $P[1]$ is assigned to $p_l = \lcp(o,v_l)$ where $o$
is the last value extracted as explained previously.  Further note that the comparisons for the first element
extracted equals its length as it's conceptual predecessor is the empty string.
So the total number of character comparisons is $|s_1| + \sum_{i=2}^M \lcp(s_{i-1},s_i)$ where
$R^{+} = s_1, s_2, \ldots s_M$ is the sequence of $M$ strings extracted from the heap over the course of
the list merge, i.e. the output list if duplicates were not removed.
This expression clearly shows that the time spent comparing strings in a string heap is a function of the
consecutive similarity of the strings in the final list, and immediately proves the expected time complexity
claim for the Uniform Scenario as the average $\lcp$ value is $O(\log_\Sigma M)$ in this scenario.

\newcommand{\LFT}{\mbox{\it LFT}}
\newcommand{\RGT}{\mbox{\it RGT}}

\section{The Collision Heap}

One might think that when merging sorted string lists that themselves have no duplicates, that there would
be in expectation very few duplicates between the lists.  This would be correct for the Uniform Scenario.
But this is not true, for instance, when the problem is to merge lists of $k$-mers generated from a shotgun data set.  
To wit, in a coverage $c$, say 40X, data set, every part of the underlying target sequence/genome has
been sampled on average 40 times and so we expect non-erroneous $k$-mers from unique parts of the target to
occur on average 40 times, and a multiple thereof if from repetitive regions.  So if one were to partition the data
into $T$ equal sized parts, sort the $k$-mers in each part, and then merge those lists, one quite often
sees the same $k$-mer in different lists.  More precisely, the chance that a given $k$-mer that occurs $c$ times
in the data set is not in a given input queue is $(1-1/T)^c$, so we expect the $k$-mer to be in $\bar e = T(1-(1-1/T)^c) \approx
T(1-e^{-c/T})$ of the input lists.  So if $T$ is say 10, then a non-erroneous, unique $k$-mer will be found in
$\bar e = 6.5$, $8.8$, $9.6$, or $9.85$ of the lists if $c = 10$, 20, 30, or 40, respectively.  It was this specific
use-case, that we call the ``{\it Shorgun Scenario}", that motivated the development of a collision heap.

The idea behind a collision heap is also very simple, namely, for each node also record whether the value at
that node is equal to its left child and its right child with a pair of boolean flags in auxiliary (bit) arrays $L[1..T]$
and $R[1..T]$.
Formally, $L[i]$ has the value of the predicate $V[H[i]] = V[H[2i]]$ and $R[i]$ has the value
of the predicate $V[H[i]] = V[H[2i+1]]$.
Again the interesting and somewhat complex part of this extension is maintaining these values during the induction
of {\it Heapify} and using them to accelerate the handling of duplicate entries.

\begin{figure}
\scriptsize
\begin{verbatim}
int     H{1..T]                                                        int G[1..T]
value   V[1..T]
boolean L[1..T]                                                        int cohort(int c, int len)
boolean R[1..T]                                                        { if (R[c]) 
                                                                           len = cohort(2c+1,len)
static void Heapify(int i, value x, int t)                               if (L[c])
{ V[t] = x                                                                 len = cohort(2c,len)
  c    = i                                                               len += 1
  while ((l = (2c)) <= T)                                                G[len] = c
    { hl = H[l]                                                          return len
      vl = V[hl]                                                       }
      if (l >= T)
        vr = INFINITY
      else   
        { hr = H[l+1]
          vr = V[hr]
        }
      if (vr > vl)
        { if (x > vl)             # Case 1L
            { H[c] = hl
              L[c] = L[l] or R[l]
              R[c] = false
              c    = l
            }
          else if (x == vl)       # Case 2L
            { H[c] = t
              L[c] = true
              R[c] = false
              return
            }
           else                    # Case 3L
            break
        }
      else if (vr < vl)
        { # Cases 1R, 2R, 3R
          . . . 
        }
      else 
        { if (x > vl)            # Case 4
            { H[c] = hl
              L[c] = L[l] or R[l]                                      domain_list S[1..T]
              R[c] = true
              c    = l                                                 1. for t = T downto 1 do
            }                                                          2.   Heapify(t,Pop(S[t]),t)
          else if (x < vl)       # Case 3
            break                                                      3. while (t = H[1]) > 0 do
          else                   # Case 5                              4.   { output V[t]
            { H[c] = t                                                 5.     len = cohort(1,0)
              L[c] = R[c] = true                                       6.     for k = 1 to len do
              return                                                   7.       { i = G[k]
            }                                                          8.         t = H[i]
        }                                                              9.         if Empty(S[t]) then
    }                                                                  10.           Heapify(i,INFINITY,0)
  H[c] = t                                                             11.         else
  L[c] = R[c] = false                                                  12.           Heapify(i,Pop(S[t]),t)
  return                                                                        }
}                                                                           }
\end{verbatim}
\normalsize
\caption{The {\it Heapify} (left) and {\it cohort} (upper right) and top-level merge (lower right) algorithms for a collision heap.
In the main algorithm {\it cohort(1,0)} identifies all of the equal next elements to be output in Line 5, and then Lines 6-12 carefully
replace each of these with its list successor.}
\end{figure}

The inductive invariant for the loop of {\it Heapify} is basically that all values are correct or {\it will be correct}
once the algorithm is complete, save for $H[c]$, $L[c]$ and $R[c]$ which need to be determined depending
on the relative values of $x$, conceptually at $c$, and those of its current children.  Let $v_l = H[V[2c]]$ and $v_r = H[V[2c+1]]$
be the strings currently at the left and right children of $c$.  There are 9 cases depending on
the relative magnitudes of $x$, $v_l$, and $v_r$, where the three that entail the condition $v_r > v_l$ are treated
by symmetry:

\vspace{6pt}
{\bf Case 1:} $v_l < v_r$ and $x > v_l$:

\hspace*{12pt}
\begin{minipage}{.9\textwidth}
By the conditions, $v_l$ will move to node $c$ and the path followed descends to $2c$.
$v_l > v_r$ implies that $R[c]$ should be {\it false}.  However, $v_l > x$ does not imply the
same for the new value of $L[c]$ as $v_l$ could be equal to the element at $2(2c)$ or $2(2c)+1$ or both
and if so, then those elements are also less than $x$ implying one or the other will replace $x$ at
$2c$.  Therefore $L[c]$ should be {\it true} as it will be correct and remain correct after the next loop iteration.  So to
recapitulate, if $L[2c]$ or $R[2c]$ are true then $L[c]$ should be set to true otherwise it should be false.
\end{minipage}

\vspace{6pt}
{\bf Case 2:} $v_l < v_r$ and $x = v_l$:

\hspace*{12pt}
\begin{minipage}{.9\textwidth}
By the conditions, the loop will terminate with $x$ finaly resting at node $c$.  By the
case conditions it is then clear that $L[c]$ is true and $R[c]$ is false.
\end{minipage}

\vspace{6pt}
{\bf Case 3:} $v_l \le v_r$ and $x < v_l$:

\hspace*{12pt}
\begin{minipage}{.9\textwidth}
Again a very simple case where the loop stops and clearly $L[c] = R[c] = \mbox{\it false}$.
\end{minipage}

\vspace{6pt}
{\bf Case 4:} $v_l = v_r$ and $x > v_l$:

\hspace*{12pt}
\begin{minipage}{.9\textwidth}
In this case, $v_l$ moves up to occupy $c$ and $x$ moves down to node $2c$.
Clearly $R[c]$ should then be true as $v_l = v_r$.
As argued in Case 1, if $v_l$ equals either of its children then the value of $L[c]$
needs to be true as one of these children is smaller than $x$.  Otherwise $L[c]$
should be false.
\end{minipage}

\vspace{6pt}
{\bf Case 5:} $v_l = v_r$ and $x = v_l$:

\hspace*{12pt}
\begin{minipage}{.9\textwidth}
Then the loop terminates and both $L[c]$ and $R[c]$ are true.
\end{minipage}

\vspace{12pt}
Figure 3 presents the complete algorithm for the collision version of {\it Heapify} embodying
the case analysis above so that the $L$ and $R$ array values are correctly maintained when
a heap update occurs.  The code is further obviously $O(\log T)$.

The value of the additional $L$ and $R$ flags is that when the top element, say $x$, is about to be extracted
as the current minimum in the heap, one can find all the additional elements equal to $x$ by recursively
visiting the children that are marked as equal according to the relevant $L$ and $R$ flags.
In Figure 3, the routine {\it PopHeap} calls the recursive routine {\it cohort} that makes a post order traversal
of the subtree of the heap of all elements equal to $x$, and places the indices of these nodes
in post order in an array $G$, returning how many of them there are.  Thus after calling {\it PopHeap}
the array $G[1..PopHeap()]$ contains the next group of equal elements.
The routine clearly takes time proportional to the number of equal elements found.

While the flags allow us to easily identify the next cohort of equal elements to extract from the
heap, there remains the somewhat more subtle problem of replacing all of them with their list
successors.  Lines 6-12 of the psuedo-code for the top-level merge at the bottom right of Figure 2
details how this is done.  Because the nodes in the cohort $G$ are in post-order, calling {\it Heapify}
on each listed node in that order guarantees a proper heap after all the elements have been replaced.
In terms of complexity suppose $e$ nodes are in the cohort for a given iteration of the loop.
While the time taken in Lines 6-12 is certainly $O(e \log T)$ we can bound this more tightly
by observing that the most time is taken when the $e$ nodes form a complete binary subtree of
the heap, that is, every node has the highest height possible.  In this case the lowest nodes are at height
$\log T - \log e$ and the sum over all $e$ nodes is dominated by this as the sum telescopes (e.g. as for
the time analysis for establishing the heap in Lines 1 and 2).  Thus the time taken is more
accurately $O(e \log (T/e))$.  Observe that when $e = T$ the time is $O(e)$ and when $e = 1$ the
time taken is $O(\log T)$.

Looking at the overall time to produce the final list $R$ where $r_i$ occurrs in $e_i$ of the lists, the
total time is $O( \sum_{i=1}^N e_i (\log T - \log e_i))$.  By the convexity of the $\log$-function $\sum_i e_i \log_2 e_i \ge N \bar e \log_2 \bar e$
where $\bar e = \sum_i e_i/N$ is the average value of $e_i$.  It thus follows that the total time
is $O( \sum_i e_i \log T  - N \bar e \log \bar e)$ = $O(M\log (T/ \bar e))$.  So when $\bar e = 1$, i.e. every
input element is unique, then the time is $O(M\log T)$ as usual.  But this gradually decreases as $\bar e$
approaches $T$ where upon the time is $O(M)$.

\section{The String Collision Heap}

Observing that the idea of a string heap and a collision heap are independent, one can combine the ideas obtaining
an $O(M \log (T/ \bar e) + S)$ time algorithm.  Further observe that the $L$- and $R$-arrays are not necessarily needed
as $L[c]$ is the same as the predicate $V[H[x]][L[x]] = \$$ where $x = 2c$ and $R[c]$ is similarly $V[H[x]][L[x]] = \$$ where
$x = 2c+1$.  In words,
the string of a child equals the string of its' parent iff the character at its' $\lcp$-value is the end of its' string.  If one has
the length $Len[t]$ of the current string from the $t^{th}$ input list, then the test is even more simply, $L[x] = Len[H[x]]$ where
$x$ is either $2c$ or $2c+1$.

\section{A Trie-Based Priority Queue for Strings}

We briefly review trie-based implementations of a string priority queue in order to explain which approach we chose to compare
against the modified heap algorithms of this paper.
Given a basic Fredkin trie, adding a new string is a matter of following the path from the root of the trie spelling the common prefix
with the new string, until its remaining suffix diverges at some node $x$.  A new out edge labelled with the first character of the remaining
suffix is added to node $x$ and trie nodes for the suffix are linked in.  Finding the minimum string in a trie is simply a matter of following
the out edge with the smallest character from each node.  To delete this minimum, one observes the last node along the minimum path
that has out degree greater than one, and then removes the minimum out edge from this divergent node and the suffix that follows.

If the out edges of each node are realized with a van Emde Boas priority queue for which add and delete are $O(\log \log \Sigma)$ and
finding the minimum is $O(1)$ then adding and deleting from the queue are both $O(\log \log \Sigma + s)$ where $s$ is the length of
the string being added or deleted.  Finding the minimum element is $O(s)$.  This gives the $O(M \log \log \Sigma + S)$ bound for the
entire merge.  If one further realizes a compact trie, wherein all nodes with out degree 1 are collapsed into their successor so that nodes
are now labeled with string fragments, the trie is guaranteed to have $O(T)$ nodes and thus the space requirement for the trie is
$O(T\Sigma)$ (excluding the space for the strings themselves).

Empirically we found that for typical values of $\Sigma$ it is actually more efficient to simply realize the out edge PQ with a $\Sigma$ element
array that is directly indexed with a character.  In addition, one keeps the current out-degree of the node and the current minimum out-edge.
With this information finding the minimum and adding a new string is just $O(s)$.  Deletion however does require traversing the out-edge
array at the divergent node looking for the new minimum out-edge and so is $O(\Sigma + s)$.  Offsetting this is the fact that the number of
string deleted/extracted from the trie is $N$ and not $M$, so the total complexity for this simple implementation is $O(N\Sigma+S)$ and
as will be seen this empirically give very good performance for the Shotgun Scenario.
 
\begin{table}[h]
\begin{center}
\footnotesize--
\begin{tabular}{ l || r || r || r || r || r || r }
\multicolumn{1}{c ||}{} & \multicolumn{1}{c ||}{} & 
   \multicolumn{1}{c ||}{Time (in sec.)} & \multicolumn{1}{c ||}{Time (in sec.)} & \multicolumn{1}{c ||}{Time (in sec.)} &
   \multicolumn{1}{c ||}{Time (in sec.)} & \multicolumn{1}{c}{Time (in sec.)} \\
\multicolumn{1}{c ||}{M, $\Sigma$, $\bar{\lcp}$, $\bar e$} & \multicolumn{1}{c ||}{T} &
  \multicolumn{1}{c ||}{Heap} & \multicolumn{1}{c ||}{String Heap} & \multicolumn{1}{c ||}{Collision Heap} &
 \multicolumn{1}{c ||}{Combination Heap}&  \multicolumn{1}{c}{Trie}  \\[3pt]
\hline
& & & & & & \\[-6pt]
 10M, 4, 10.8, 1.000 & 4 & .654 & {\bf .506} & .597 & .542 & .984 \\
 & 8 & .848 & {\bf .633} & .851 & .673 & 1.022 \\
 & 16 & 1.057 & {\bf .776} & 1.130 & .810 & 1.060 \\
 & 32 & 1.268 & {\bf .907} & 1.407 & .950 & 1.097 \\
 & 64 & 1.479 & {\bf 1.053} & 1.658 & 1.090 & 1.144 \\
 & 128 & 1.650 & {\bf 1.183} & 1.914 & 1.230 & 1.249 \\
 & 256 & 1.897 & {\bf 1.323} & 2.201 & 1.373 & {\bf 1.331} \\
 \hline
& & & & & & \\[-6pt]
100M, 4, 12.5, 1.000 & 4 & 7.04 & {\bf 5.21} & 6.29 & 5.54 & 9.99 \\
  & 8 & 8.99 & {\bf 6.47} & 8.96 & 6.85 & 10.81 \\
  & 16 & 11.18 & {\bf 7.88} & 11.84 & 8.12 & 11.10 \\
  & 32 & 13.57 & {\bf 9.15} & 14.75 & 9.60 & 11.50 \\
  & 64 & 16.16 & {\bf 10.58} & 17.86 & 10.90 & 12.01 \\
  & 128 & 18.69 & {\bf 12.05} & 21.18 & 12.34 & 13.09 \\
  & 256 & 21.63 & {\bf 13.64} & 24.53 & 14.00 & 13.77 \\
 \hline
& & & & & & \\[-6pt]
1000M, 4, 14.1, 1.000 & 4 & 72.9 & {\bf 52.5} & 66.0 & 55.7 & 99.7 \\
  & 8 & 94.7 & {\bf 65.5} & 95.2 & 68.9 & 107.9 \\
  & 16 & 115.9 & {\bf 77.7} & 121.9 & 81.0 & 111.6 \\
  & 32 & 140.2 & {\bf 91.1} & 154.6 & 95.6 & 116.2 \\
  & 64 & 168.5 & {\bf 106.6} & 187.4 & 111.2 & 123.4 \\
  & 128 & 195.5 & {\bf 121.1} & 221.8 & 125.7 & 132.2 \\
  & 256 & 229.5 & {\bf 139.2} & 265.3 & 142.9 & 143.2 \\
 \hline
& & & & & & \\[-6pt]
1000M, 8, 9.3, 1.000 & 4 & 61.4 & {\bf 51.0} & 57.2 & 53.8 & 89.7 \\
 & 8 & 77.7 & {\bf 64.3} & 81.1 & 67.3 & 93.6 \\
 & 16 & 94.2 & {\bf 76.7} & 103.0 & 80.0 & 95.7 \\
 & 32 & 108.5 & {\bf 90.5} & 126.6 & 94.1 & 98.3 \\
 & 64 & 126.4 & 105.4 & 149.3 & 110.2 & {\bf 102.4} \\
 & 128 & 145.9 & 118.9 & 174.1 & 123.9 & {\bf 110.2} \\
 & 256 & 168.5 & 133.5 & 202.4 & 137.8 & {\bf 113.9} \\
 \hline
& & & & & & \\[-6pt]
1000M, 16, 6.8, 1.000 & 4 & 56.1 & {\bf 50.0} & 53.4 & 50.8 & 85.2 \\
  & 8 & 69.2 & {\bf 60.2} & 71.1 & 63.9 & 85.0 \\
  & 16 & 81.4 & {\bf 74.1} & 92.0 & 75.2 & 87.7 \\
  & 32 & 95.1 & {\bf 88.1} & 110.5 & 89.2 & {\bf 88.5} \\
  & 64 & 109.6 & 99.7 & 132.7 & 103.9 & {\bf 91.3} \\
  & 128 & 126.7 & 113.6 & 157.0 & 119.1 & {\bf 100.2} \\
  & 256 & 150.0 & 129.3 & 181.0 & 134.6 & {\bf 101.7}
 \end{tabular}
\end{center}
\caption{Performance for the Uniform Scenario.}
\label{default}
\end{table}

\section{Empirical Performance}

We implemented string list merging programs using a regular heap {\tt Heap}, a string heap {\tt Sheap}, a collision heap
{\tt Cheap}, a string-collision heap {\tt SCheap}, and a simple, compact trie {\tt Trie} and measured their performance on a 2019 Mac Pro with
a 2.3 GHz Intel Core i9 processor, 64GB of memory, and 8TB of SSD disk.  All the codes are available at GitHub at
the url {\tt github.com/thegenemyers/HEAPS}.

In the first set of timing experiments over synthetic data, for a given setting of parameters $N$, $T$, and $\Sigma$ we generated
$T$ input files, each with $N$ $20$-mers where every $20$-mer over a $\Sigma$ character ASCI alphabet occurs with equally likelihood,
that is, the Uniform Scenario introduced in the introduction.
For such data we expect the $\lcp$ between successive elements in the output list to be on average $\log_\Sigma M$ and $\bar e$
to be 1 given that the average $\lcp$ is less than $20$ for all trials considered.  In Table 1, we present timings where $\Sigma$ was set
to 4, $T$ was set from 4 to 256 in steps of 2x, and $M$ was set to $10^x$ for x = 7, 8, and 9.
In addition, for $M = 10^9$, we also generated data sets where $\Sigma$ was also set to 8 and 16 to see the dependence of
the programs, especially $Trie$, on $\Sigma$.

The timings confirm that the all algorithms are linear in $M$ and the heap algorithm linear in $\log T$.
As $M$ becomes larger or $\Sigma$ becomes smaller the average $\lcp$
between consecutive strings increases and so as expected the string heap becomes progressively faster than a regular heap.
The combination heap tracks the performance of the string heap but lags by about 5\% for all parameter values due to the
additional overhead of maintaining information about collisions which in these experiments are basically do not occur.
The tries behavior is basically constant edging up slightly with $T$ due to an increasing in the branching layers in the prefix of the trie.
Counter intuitively, the trie becomes faster with larger $\Sigma$ as this reduces the expected number of branching layers in the trie which
dominates the minor cost of searching for the smallest out-edge of a single node when deleting an entry.
Thus, ultimately as $T$ increases the trie becomes the fastest, at 256 for $\Sigma = 4$, 64 for $\Sigma = 8$, and 32 for $\Sigma = 16$.

In the second set of timing experiments, we produced synthetic data sets of $K$-mers where they followed the
Shotgun Scenario.  We fixed $M$ at $10^9$, $K$ at 20, and then for each of $T = 4$, $8$, and $16$, we varied $C$ such
that $C/T$ = $.25$, $.5$, $1.0$, $1.5$, $2.0$, and $3.0$.

\begin{table}[h]
\begin{center}
\footnotesize--
\begin{tabular}{ r || r || r || r || r || r || r  || r || r}
  &  &  &  &
   \multicolumn{1}{c ||}{Time (in sec.)} & \multicolumn{1}{c ||}{Time (in sec.)} & \multicolumn{1}{c ||}{Time (in sec.)} &
   \multicolumn{1}{c ||}{Time (in sec.)}  &  \multicolumn{1}{c}{Time (in sec.)} \\
\multicolumn{1}{c ||}{T} & \multicolumn{1}{c ||}{C} & \multicolumn{1}{c ||}{$\bar{\lcp}$} & \multicolumn{1}{c ||}{$\bar e$} & 
  \multicolumn{1}{c ||}{Heap} & \multicolumn{1}{c ||}{String Heap} & \multicolumn{1}{c ||}{Collision Heap} &
  \multicolumn{1}{c ||}{Combined Heap} &  \multicolumn{1}{c}{Trie} \\[3pt]
\hline
 & & & & & & & \\[-6pt]
 4 & 0 & 14.1 & 1.0 & 70.6 & {\bf 54.2} & 65.8 & 58.0 & 98.4 \\
    & 1 & 15.7 & 1.4 & 73.7 & {\bf 53.6} & 59.8 & 56.3 & 84.6 \\
    & 2 & 16.8 & 2.0 & 69.7 & {\bf 50.1} & 49.9 & 50.7 & 66.5 \\
    & 4 & 17.7 & 2.8 & 63.2 & {\bf 42.6} & 38.6 & 43.3 & 48.0 \\
    & 6 & 18.0 & 3.3 & 60.2 & {\bf 38.4} & 34.0 & 38.7 & 41.0 \\
    & 8 & 18.1 & 3.6 & 58.5 & 35.8 & 31.7 & {\bf 35.1} & 36.5 \\
    & 12 & 18.2 & 3.9 & 57.0 & 34.0 & 29.8 & {\bf 32.8} & 33.9 \\
\hline
  & & & & & & & \\[-6pt]
 8 & 0 & 14.1 & 1.0 &   95.0 & {\bf 66.4} & 96.7 & 70.1 & 113.5 \\
    & 2 & 17.0 & 2.1 &   93.6 & {\bf 64.7} & 79.1 & 65.8 & 79.8 \\
    & 4 & 18.1 & 3.5 &   88.4 & 57.8 & 60.9 & {\bf 56.0} & 61.1 \\
    & 8 & 18.7 & 5.3 &   82.5 & 48.1 & 45.9 & 46.9 & {\bf 45.9} \\
    & 12 & 18.9 & 6.4 &  79.1 & 42.5 & 38.8 & 40.6 & {\bf 39.6} \\
    & 16 & 19.0 & 7.1 &   77.9 & 39.3 & 34.9 & 37.0 & {\bf 35.9} \\
    & 24 & 19.0 & 7.7 &   76.9 & 35.4 & 31.7 & {\bf 33.2} & 33.5 \\
\hline
 & & & & & & & \\[-6pt]
16 & 0 & 14.1 & 1.0 &   117.9 & {\bf 79.3} & 124.6 & 82.8 & 113.7 \\
     & 4 & 18.2 & 3.9 &   109.8 & 73.0 & 85.6 & 70.7 & {\bf 63.0} \\
     & 8 & 18.9 & 6.6 &   101.3 & 63.1 & 65.1 & 59.3 & {\bf 50.3} \\
     & 16 & 19.3 & 10.4 &   97.3 & 54.3 & 48.8 & 47.5 & {\bf 40.1} \\
     & 24 & 19.4 & 12.6 &   94.9 & 48.1 & 41.7 & 40.9 & {\bf 34.9} \\
     & 32 & 19.4 & 14.0 &   93.8 & 44.0 & 37.9 & 36.8 & {\bf 33.2} \\
     & 48 & 19.5 & 15.3 &   91.8 & 39.1 & 34.1 & 33.5 & {\bf 30.6}
\end{tabular}
\end{center}
\caption{Performance for the Shotgun Scenario.}
\label{default}
\end{table}

As the number of collisions increases, the collision heap eventually overtakes the string heap, with again the combination heap
tracking the better of the two with an overhead of 5\% or so.
But our trie implementation quickly becomes faster for larger values of $T$ due to its $O(N\Sigma+S)$ complexity.  The collision heaps, in terms
of $N$, has complexity $O(N \bar e \log (T/ \bar e))$ which explains the behavior.  Basically, the number of elements in the trie
decreases rapidly from $T$ toward $1$ as collisions occur greatly accelerating its operation.
Nonetheless, the table reveals that for smaller values of $T$ the heap algorithms are competitive or even superior.

The final set of experiments were for $k$-mers from a real shotgun sequencing data set, the motivating example for this work.
For high-accuracy read data sets $k$ is typically chosen at 40 or more.  The other difference with the synthetic Shotgun Scenario
is that the $k$-mers occur with a complex frequency profile wherein some $k$-mers occur with frequency about $C$, but for example,
80\% of the $k$-mers contain errors and occur once, others from a haplotype region occur roughly $C/2$ times, and so on.
So in Table 3 below one will see that $\bar e$ is significantly less than for the synthetic examples, yet still substantially elevated.
Interesting in these cases the collision heap or combined heap perform best because they respond continuously to the collisions, beating
out the string heap, and the $\lcp$ is very near $k$, thus beating out the trie.

\begin{table}[h]
\begin{center}
\footnotesize--
\begin{tabular}{ r || r || r || r || r || r || r || r  || r || r}
 &  &  &  & &
   \multicolumn{1}{c ||}{Time (in sec.)} & \multicolumn{1}{c ||}{Time (in sec.)} & \multicolumn{1}{c ||}{Time (in sec.)} &
   \multicolumn{1}{c ||}{Time (in sec.)}  &  \multicolumn{1}{c}{Time (in sec.)} \\
\multicolumn{1}{c ||}{M} & \multicolumn{1}{c ||}{T} & \multicolumn{1}{c ||}{C} & \multicolumn{1}{c ||}{$\bar{\lcp}$} &
\multicolumn{1}{c ||}{$\bar e$} & 
  \multicolumn{1}{c ||}{Heap} & \multicolumn{1}{c ||}{String Heap} & \multicolumn{1}{c ||}{Collision Heap} &
  \multicolumn{1}{c ||}{Combined Heap} &  \multicolumn{1}{c}{Trie} \\[3pt]
\hline
 & & & & & & & & & \\[-6pt]
 333M & 4 & 4 & 28.3 & 2.1 &  33.2 & 23.1 & {\bf 22.4} & 24.8 & 29.1 \\
 666M& 8 & 8 & 32.9 & 3.4 &  85.7 & 56.1 & {\bf 55.0} & {\bf 55.0} & 60.5 \\
 1333M & 16 & 16 & 35.4 & 4.9 &  210.2 & 123.3 & 118.5 & {\bf 114.7} & 121.8 \\
 \hline
  & & & & & & & & & \\[-6pt]
 484M & 4 & 8 & 30.3 & 2.5 &  50.7 & 32.5 & {\bf 29.9} & 33.7 & 38.3 \\
 968M & 8 & 16 & 33.6 & 3.6 &  127.4 & 74.9 & {\bf 71.3} & 73.6 & 84.0 \\
 1936M & 16 & 32 & 35.4 & 4.6 &  314.9 & 175.0 & 164.7 & {\bf 159.2} & 183.4 \\
 \hline
 & & & & & & & & & \\[-6pt]
566M & 4 & 12  & 30.4 & 2.4 & 59.6 & 37.6 & {\bf 35.2} & 39.2 & 44.0 \\
1232M & 8 & 24  & 33.3 & 3.3 & 149.9 & 87.1 & {\bf 84.3} & 86.3 & 98.5 \\
 2464M & 16 & 48 & 34.8 & 4.0 & 382.2 & 212.0 & 205.3 & {\bf 194.7} & 230.4 \\
\end{tabular}
\end{center}
\caption{Performance on real sequencing data with $k = 40$.}
\label{default}
\end{table}                                   

In summary, the string heap performs best when the average $\lcp$ value increases, and the collision ratio is low.
The collision heap always performs better than the string heap when collisions become high.
The combination heap tracks the better of the two combined methods, lagging by about 5\%.
The trie data structure is generally the best for large $T$ or pure collision scenarios, but on real high-fidelity shotgun 
data sets the collision and combination heaps proved superior.

\section{Acknowledgments}
I would like to acknowledge Richard Durbin and Travis Gagie, for their encouragement to write up this work, Shane McCarthy
for providing the data sets that made the need for a collision heap apparent, Gonzalo Navarro for suggesting the trie approach
and pointing at Thorup's work, and Shinichi Morishita and Yoshihiko Suzuki
for their many helpful comments and review of the work.


\begin{thebibliography}{1}

\bibitem{Knuth}
Knuth, D. (1998)
Chapter 5.4.1: Merging and Replacement Selection, in
The Art of Computer Programming Vol. 3
(2nd edition, Addison Wesley), 252,255.

\bibitem{VGP}
Rhie, A., McCarthy, S., Frederigo, O., ..., Myers, E.W., Durbin, R., Phillippy, A.M., and Jarvis, E. (2021)
Towards Complete and Error-Free Genome Assemblies of all Vertebrate Species.
{\it Nature} {\bf 592}, 737-746.

\bibitem{KMC3}
Kokot, M., Dlugosz, M., and Deorowicz, S. (2017)
KMC3: Counting and Manipulating $k$-mer Statistics.
{\it Bioinformatics} {\bf 33}, 2759-2761.

\bibitem{PBJelly}
Marcais, G. and Kingsford, C. (2011)
A Fast, Lock-Free Approach for Efficient Parallel Counting of Occurrences of $k$-mers.
{\it Bioinformatics} {\bf 27}, 764-770.

\bibitem{FastK}
Myers, E. (2020)
$\mathtt{https://github.com/thegenemyers/FASTK}$.

\bibitem{Corman}
Cormen, T.H., Leiserson, C.E., Rivest, R.L., and Stein, C.  (2009)
Introduction to Algorithms (3rd. edition, MIT Press), 151-169.

\bibitem{Thorup}
Thorup, M. (2000)
On RAM Priority Queues.
{\it SIAM J. on Computing} {\bf 30}, 86-109.

\bibitem{Fredkin}
Fredkin, E. (1960)
Trie Memory.
{\it Comm. of the ACM} {\bf 3}, 490-499.

\bibitem{Boas}
van Emde Boas, P., Kaas, R., and Zijlstra, E. (1977)
Design and Implementation of an Efficient Priority Queue.
{\it Mathematical Systems Theory} {\bf 10}, 99-127.

\end{thebibliography}
\end{document}